\documentclass[runningheads,a4paper]{llncs}

\usepackage{amssymb}
\usepackage{graphicx}

\usepackage{url}
\urldef{\mailsa}\path|{nikolai.antonov}@pobox.spbu.ru|
\urldef{\mailsb}\path|{kollya12}@mail.ru|
\newcommand{\keywords}[1]{\par\addvspace\baselineskip
\noindent\keywordname\enspace\ignorespaces#1}

\begin{document}

\mainmatter  

\title{Two-Loop Calculation of the Anomalous Exponents
in the Kazantsev--Kraichnan Model of Magnetic Hydrodynamics}

\titlerunning{Calculation of the Anomalous Exponents in the Kraichnan MHD Model}

%
%
\author{N.V. Antonov%
\and N.M. Gulitskiy}
\authorrunning{N.V. Antonov \and N.M. Gulitskiy}

\institute{Department of Physics, St. Petersburg State University, \\
Ulyanovskaya 3, Petrodvorets, St. Petersburg 198504 Russian Federation;\\ 
D.I.~Mendeleyev Institute for Metrology,\\
Moskovsky pr. 19, St.Petersburg, 190005, Russian Federation,\\
\mailsa\\
\mailsb\\
\url{http://hep.niif.spbu.ru}
}

%
%

\toctitle{Calculation of the Anomalous Exponents in the Kraichnan MHD Model}
\maketitle

\begin{abstract}
The problem of anomalous scaling in magnetohydrodynamics turbulence is
considered within the framework of the kinematic approximation, in the
presence of a large-scale background magnetic field. Field theoretic
renormalization group methods are applied to the
Ka\-zan\-t\-sev--Kraichnan model of a passive vector advected by the
Gaussian velocity field with zero mean and correlation function
$\propto\delta(t-t')/k^{d+\epsilon}$. Inertial-range anomalous scaling
for the tensor pair correlators is established as a consequence of the
existence in the corresponding operator product expansions of
certain ``dangerous'' composite operators, whose negative critical
dimensions determine the anomalous exponents. The main technical result
is the
calculation of the anomalous exponents in the order $\epsilon^2$ of the
$\epsilon$ expansion (two-loop approximation).
\end{abstract}

\section{Introduction}

Much attention has been paid recently to a simple model of the passive
advection of a scalar quantity by a Gaussian short-correlated velocity
field, introduced first by Obukhov \cite{Obukhov} and Kraichnan
\cite{Kraichnan-1968}. The structure functions of the field in this
model exhibit anomalous scaling behavior, and the corresponding
anomalous exponents can be calculated within regular expansions
in a small parameter.

Effects of intermittency and anomalous scaling are even more important
for vector fields. In particular, the large-scale intense anisotropic
magnetic fields coexist with small-scale turbulent activity in solar wind,
see e.g. \cite{Salem} and references therein.

In this communication, we discuss the anomalous scaling of magnetic fields
in the presence of large-scale anisotropy within the framework of the
kinematic Kazantsev-Kraichnan model, using the field theoretic methods of
renormalization group and operator product expansion. We extend the
one-loop results derived in \cite{Antonov-Lanotte-Mazzino}
to the two-loop order of the $\epsilon$-expansion.

\subsection{Kinematic MHD Kazantsev-Kraichnan Model}

In the presence of a mean component \mbox{\boldmath $\theta^0$}
(actually supposed to be varying on a very large scale $L$, the largest
one in our problem) the kinematic MHD equations, describing the evolution
of the fluctuating part
\mbox{\boldmath $\theta$}=\mbox{\boldmath $\theta$}$(x)$
of the magnetic field, are \cite{MagneticFields}
\begin{equation}
\label{stochastic-mgd}
\partial_{t}\theta_{i}+{\bf v}\cdot{\mbox{\boldmath $\partial$}}\theta_{i}=\mbox{\boldmath $\theta$}\cdot{\mbox{\boldmath $\partial$}}v_{i}+\mbox{\boldmath $\theta^0$}\cdot{\mbox{\boldmath $\partial$}}v_{i}+\nu_0\partial^2\theta_{i}, \qquad i=1,\dots,d,
\end{equation}
where the term $\mbox{\boldmath $\theta^0$}\cdot{\mbox{\boldmath $\partial$}}v_{i}\equiv f_i$ effectively plays the same role as an external force, driving the system, with correlator 
\begin{equation}
\label{corr-f-Intro}
\left\langle f_i(x)f_j(x')\right\rangle=\delta(t-t')C_{ij}(mr),
\end{equation}
where $C$ is some function finite at $r=0$ and decaying for $r\to\infty$
and $m=1/L$ is the reciprocal of the integral turbulence scale.
Here and below
$x\equiv\left\{t,{\bf x}\right\},$
${\mbox{\boldmath $\partial$}}\equiv\left\{\partial_i=\partial /
\partial {x_i}\right\},$
$\partial^2\equiv\partial_i\partial_i\equiv\Delta$ is the Laplace operator,
$d$ is the dimensionality of $\bf x$ space,
${\bf v}(x)$ is the velocity field. Both $\bf v$ and \mbox{\boldmath $\theta$} are divergent-free (solenoidal) vector fields: $\partial_iv_i=\partial_i\theta_i=0$.

In the real problem, $\bf v$ obeys the NS equation with the additional
Lorentz force term $\propto({\mbox{\boldmath $\partial$}}\times\mbox{\boldmath
$\theta$})\times\mbox{\boldmath $\theta$},$ which describes the effects
of the magnetic field on the velocity field. The framework of our analysis
is the kinematic MHD problem, where the reaction of the magnetic field
\mbox{\boldmath $\theta$} on the velocity field is neglected. We assume
that at the initial stages \mbox{\boldmath $\theta$} is weak and does not
affect the motions of the conducting fluid: it becomes then a natural
assumption to consider the dynamics linear in the magnetic field strength.

More precisely, we shall consider a simplified model, in which ${\bf v}(x)$
is a Gaussian random field, homogeneous, isotropic and $\delta$-correlated
in time, with zero mean and covariance
\begin{equation}
\label{corr-v-Intro}
\left\langle v_i(x)v_j(x') \right\rangle =
D_0\frac{\delta(t-t')}{(2\pi)^d}\int d{\bf k}P_{ij}({\bf k})
k^{-d-\epsilon}
\cdot e^{[i{\bf k}\cdot({\bf x}-{\bf x'})]},
\end{equation}
where $P_{ij}({\bf k})=\delta_{ij}-k_i k_j/k^2$ is the transverse projector,
${\bf k}$ is the momentum, $k=\left|{\bf k}\right|,$ $D_0$ is an amplitude
factor, $d$ is dimensionality of the $\bf x$ space and $\epsilon$ is a free
parameter. The IR regularization is provided by the cutoff in the integral
from below at $k\cong m\propto1/L$. The case of anisotropic velocity
ensemble was studied in \cite{HHJMS}.


\section{Field Theoretic Formulation}

This stochastic problem is equivalent to the field theoretic model of the set of three fields $\Phi=\big\{\mbox{\boldmath $\theta,\theta', v$} \big\}$
with action functional \cite{Vasiliev}:
\begin{equation}
\label{action-fieldth}
S(\Phi)={\mbox{\boldmath $\theta'$}}D_{\theta}{\mbox{\boldmath $\theta'$}}/2+\theta'_i\left[-\partial_{t}\theta_{i}+\nu_0\Delta\theta_i-\partial_{k}(v_{k}\theta_{i}-v_{i}\theta_{k})\right]-{\mbox{\boldmath $v$}}D_{v}^{-1}{\mbox{\boldmath $v$}}/2,
\end{equation}
where the first four terms represent the De Dominicis--Janssen-type
action for the stochastic problem (\ref{stochastic-mgd}--\ref{corr-f-Intro}) at fixed $\bf v$, and the last term represents the Gaussian averaging over $\bf v$.
$D_{\theta}=\langle ff\rangle$ and
$D_{v}=\langle vv\rangle$ are the correlators (\ref{corr-f-Intro}) and (\ref{corr-v-Intro}) respectively, the
required integrations over $x=\{t,\bf{x}\}$ and summations over the vector indices are understood.

The diagrams of the perturbation theory are constructed of the four elements.
In the $\omega, \bf k$ representation the factor $i[k_a\delta_{bc}-p_b\delta_{ac}]$
corresponds to the vertex, and the lines $vv$, $\theta\theta$ and $\theta\theta'$
correspond to the bare propagators
\begin{equation}
\begin{tabular}{c c c c c l}
\parbox{0.15\textwidth}{\includegraphics[width=0.15\textwidth,clip]{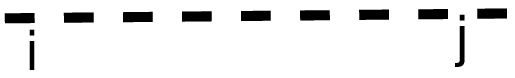}} &
\raisebox{7pt}{$=\frac{P_{ij}(\bf{k})}{k^{d+\epsilon}},$} &
\parbox{0.15\textwidth}
{\includegraphics[width=0.15\textwidth,clip]{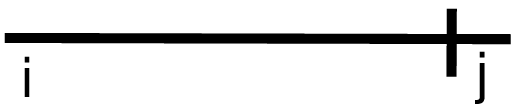}} &
\raisebox{7pt}{$=\frac{P_{ij}(\bf{k})}{-i\omega+\nu k^2},$} &
\parbox{0.15\textwidth}{\includegraphics[width=0.15\textwidth,clip]{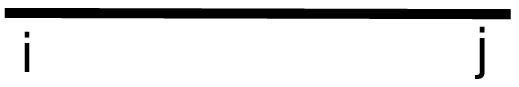}} &
\raisebox{7pt}{$=\frac{C_{ij}(\bf{k})}{\omega^2+\nu^2k^4},$} \\
\end{tabular}
\label{propagatorsInd-fieldth}
\end{equation}
where $C_{ij}(\bf {k})$ is the Fourier transform of the function
$C_{ij}$ from Eq. (\ref{corr-f-Intro}).

The UV divergences manifest themselves as poles in $\epsilon$ in the diagrams.
For the complete elimination of these divergences it is sufficient to perform the multiplicative renormalization of the parameters $\nu_0$ and $g_0$ with the only
independent renormalization constant $Z_{\nu}$ (see Ref. \cite{Adzhemyam-Antonov-1998}):
\begin{equation}
\label{RConst-fieldth}
\mbox{
\begin{tabular}{c c c}
$\nu_0=\nu Z_{\nu},$ & $g_0=g\mu^\epsilon Z_g,$ & $Z_{g}^{-1}=Z_{\nu}.$ \\
\end{tabular}}
\end{equation}

The exact response function
$G_{ij}\equiv\left\langle \theta_i\theta'_j\right\rangle$
satisfies the standard Dyson equation with just one self-energy diagram,
and therefore one can obtain an exact expression for
the renormalization constant $Z_\nu$ in the MS scheme:
\begin{equation}
\label{Znu-fieldth}
Z_{\nu}=1-u\frac{d-1}{2d}\cdot\frac{1}{\epsilon}, \quad
u = g S_{d}/(2\pi)^{d},
\end{equation}
where $S_{d}$ is the area of the unit sphere in $d$-dimensional space.

\subsection{RG Equations for Composite Operators}

The basic RG equation for a multiplicatively renormalizable quantity $F=Z_F\cdot F_R$ (correlation function, composite operator \itshape etc\normalfont) has the form
\begin{equation}
\label{RG-fieldth}
\mbox{
\begin{tabular}{c c c}
$[D_{RG}+\gamma_F]F_R=0,$ & $D_{RG}=D_{\mu}+\beta\partial_u-\gamma_{\nu}D_{\nu},$ & $D_x=x\partial/\partial x,$ \\
\end{tabular}}
\end{equation}
where RG functions $\beta$ and $\gamma$ (anomalous dimension) are defined as
\begin{equation}
\label{Def-BetaGamma-fieldth}
\mbox{
\begin{tabular}{c c c}
$\beta=\tilde{D_{\mu}}u,$ & $\gamma_F=\tilde{D_{\mu}}\ln Z_F$ & $\forall Z_F,$ \\
\end{tabular}}
\end{equation}
where $\tilde{D_{\mu}}$ is the operation $D_{\mu}$ at fixed bare parameters.

From the analysis of RG functions  it follows, that the RG equations (\ref{RG-fieldth}) possess an IR stable positive fixed point $u_*:$
\begin{equation}
\label{FixPoint-fieldth}
\mbox{
\begin{tabular}{c c c}
$u_{*}=\frac{2d}{d-1}\epsilon$, & $\beta(u_{*})=0$, & $\beta'(u_{*})>0$. \\
\end{tabular}}
\end{equation}
The value of anomalous dimension $\gamma_\nu(u)$ at fixed point $u_*$ is
\begin{equation}
\label{gamma*-fieldth}
\gamma_\nu^*\equiv\gamma_\nu(u_*)=\epsilon.
\end{equation}

This fact implies that correlation functions of this model exhibit scaling behavior; the corresponding critical dimensions $\Delta\left[F\right]\equiv\Delta_F$  can be calculated as series in $\epsilon$. For the basic fields and quantities the dimensions are found exactly \cite{Adzhemyam-1998}:
\begin{equation}
\label{CriticalDimFields-fieldth}
\mbox{
\begin{tabular}{c c c}
$\Delta_{\theta}=-1+\epsilon/2,$ & $\Delta_{\theta'}=d+1-\epsilon/2,$ &
$\Delta_{\omega}=1-\epsilon$ \\
\end{tabular}}
\end{equation}
(there is no corrections of order $\epsilon^2$ and higher, this is a consequence of the exact equality $\gamma_{\nu}(u_{*})=\epsilon$).

Let $G(r)=\left\langle F_1(x)F_2(x')\right\rangle$ be a single-time two-point
quantity; for example, the pair correlation function of the primary fields
$\Phi=\big\{\mbox{\boldmath $\theta,\theta', v$} \big\}$ or some multiplicatively renormalizable
composite operators. The solution of the RG equation gives:
\begin{equation}
\label{GAsimptotic-fieldth}
G(r)\cong\nu_0^{d^\omega_G}\Lambda^{d_G}(\Lambda r)^{-\Delta_G}\xi(mr),
\end{equation}
where the canonical dimensions $d^\omega_G$, $d_G$ and the critical dimension $\Delta_G$ of the function $G(r)$ are equal to the sums of the corresponding dimensions
of the quantities $F_i.$

This representation describes the behavior of the correlation functions for
$\Lambda r\gg1$ and any fixed value of $mr$. The inertial range
$\Lambda^{-1}=l\ll r\ll L=m^{-1}$ corresponds to the additional
condition $mr\ll1$, the form of the functions $\xi(mr)$ in the
interval $mr\ll1$ is studied using the operator product expansion (OPE).


\section{Operator Product Expansion}

According to the OPE, the single-time product $F_1(x)F_2(x')$ of two
renormalized operators has the form
\begin{equation}
\label{F-ope}
F_1(x)F_2(x')=\sum_\alpha C_\alpha({\bf r})F_\alpha({\bf x},t),
\end{equation}
where ${\bf x}\equiv{\bf (x+x')}/2=const,$ ${\bf r}\equiv{\bf x-x'}\rightarrow0,$
the functions $C_\alpha$ are the Wilson coefficients regular in $m^2$ and $F_\alpha$ are all possible renormalized local composite operators allowed by symmetry, with definite critical dimensions
$\Delta_\alpha$.

The renormalized correlator $\left\langle F_1(x)F_2(x')\right\rangle$ is
obtained by averaging Eq. (\ref{F-ope}) with the weight $\exp S_R$;
hence the desired asymptotics for the correlator\\
$\left\langle F_1(x)F_2(x')\right\rangle$ is the sum, in which the operator
possessing the minimal dimension gives the leading term:
\begin{equation}
\label{xiAsimp-OPE}
\xi(mr)\cong const\cdot(mr)^{\Delta_{min}}.
\end{equation}

The feature typical to the models describing turbulence is the existence
of composite operators with negative critical dimensions, for example,
critical dimension of the field $\theta_i$ is
$\Delta_\theta=(-1+\epsilon/2)$; their contributions in the OPE lead
to singular behavior of the scaling functions at $mr\rightarrow0$, that is,
to the anomalous scaling. The operators with minimal $\Delta_F$ are those
involving the maximal possible number of fields $\theta$ and the minimal
possible number of derivatives. Therefore the needed
operators are tensors, constructed from the fields $\theta_i$ themselves:
\begin{equation}
\label{Fnl-OPE}
\mbox{
\begin{tabular}{c c}
$F_{nl}=\theta_{i_1}\cdots\theta_{i_l}(\theta_i\theta_i)^p,$ & $n=l+2p.$ \\
\end{tabular}}
\end{equation}
The critical dimension of any multiplicatively renormalizable quantity
$F=Z_F\cdot F_R$ is
$\Delta_F=d^k_F+\Delta_\omega d^\omega_F+\gamma^*_F$. Then
for the operator $F_{nl}$ we obtain
\begin{equation}
\label{DeltaMin-OPE}
\Delta_{F_{nl}}=n(-1+\epsilon/2)+\gamma^*_{F_{nl}},
\end{equation}
and the leading asymptotic term of the correlator
$\left\langle F_{nl}(x)F_{pq}(x') \right\rangle$ in the $j$th anisotropic
sector has the form
 \begin{equation}
\label{F-Asympt-OPE}
\left\langle F_{nl}(x)F_{pq}(x') \right\rangle\propto
(\Lambda r)^{-\Delta_{F_{nl}}-\Delta_{F_{pq}}}(mr)^{\Delta_{F_{n+p,j}}}.
\end{equation}

Thus one has to calculate critical dimensions $\Delta_{F_{nl}}$ of the
operators $F_{nl}$.


\section{Scalarization of the Diagrams}

The operator $F_{nl}=\theta_{i_1}\cdots\theta_{i_l}(\theta_i\theta_i)^p,$ $n=l+2p$ is renormalized multiplicatively, $F_{nl}=Z_{nl}\cdot F_{nl}^R,$ and the renormalization constants $Z_{nl}=Z_{nl}(g,\epsilon,d)$ are determined by the requirement that the 1-irreducible correlation function
\begin{eqnarray}
\left\langle F^R_{nl}(x)\theta(x_1)\cdots\theta(x_n)\right\rangle_{1-irr}&=&Z_{nl}^{-1}\left\langle F_{nl}(x)\theta(x_1)\cdots \theta(x_n)\right\rangle_{1-irr}\nonumber\\
&\equiv &Z_{nl}^{-1}\Gamma_{nl}(x;x_1,\dots, x_n)\label{Gnl-scalar}
\end{eqnarray}
be UV finite in renormalized theory, i.e., have no poles in $\epsilon$ when expressed in renormalized variables (\ref{RConst-fieldth}).

Below we present, along with respective symmetry coefficients, all
the diagrams needed for the two-loop calculation of the function
$\Gamma_{nl}$, except for those with the self-energy insertions in the $\theta\theta'$ lines.
$$
\begin{tabular}{r c c r c c c c c}
$\Gamma^{(1)}=\frac{1}{2}$ &
\raisebox{-20pt}{
\includegraphics[width=0.12\textwidth,clip]{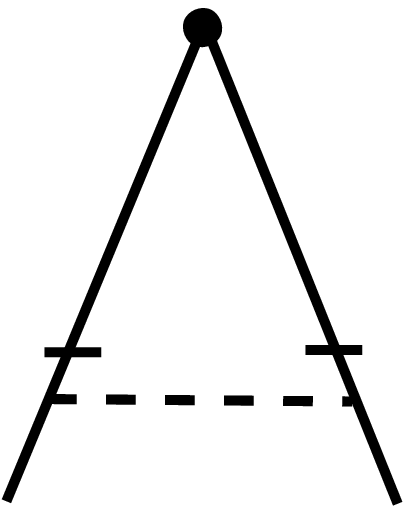}} &
, &
$\Gamma^{(2)}=\frac{1}{2}$ &
\raisebox{-20pt}{
\includegraphics[width=0.12\textwidth,clip]{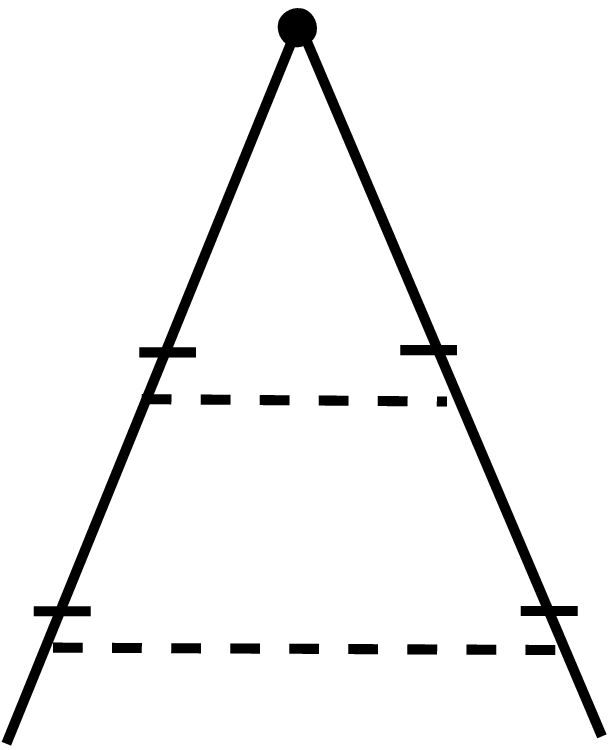}} &
$+$ &
\raisebox{-20pt}{
\includegraphics[width=0.12\textwidth,clip]{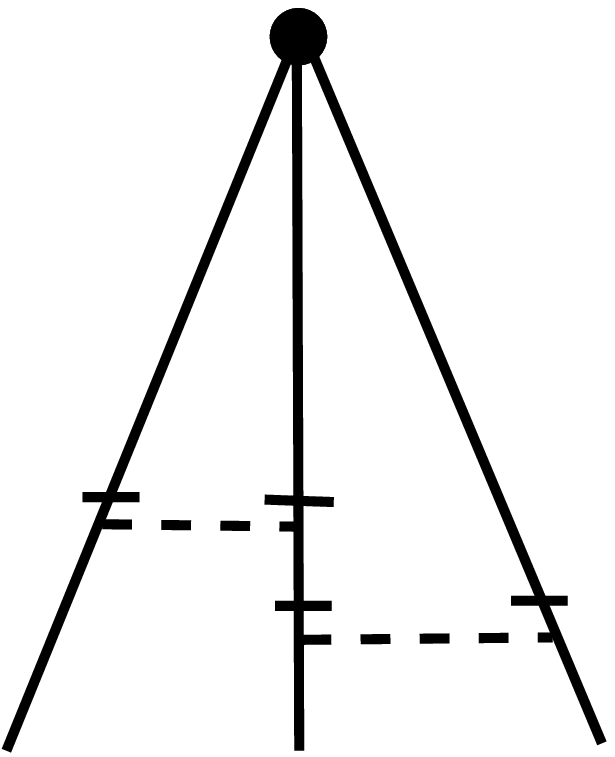}} &
$+\frac{1}{8}$ &
\raisebox{-20pt}{
\includegraphics[width=0.15\textwidth,clip]{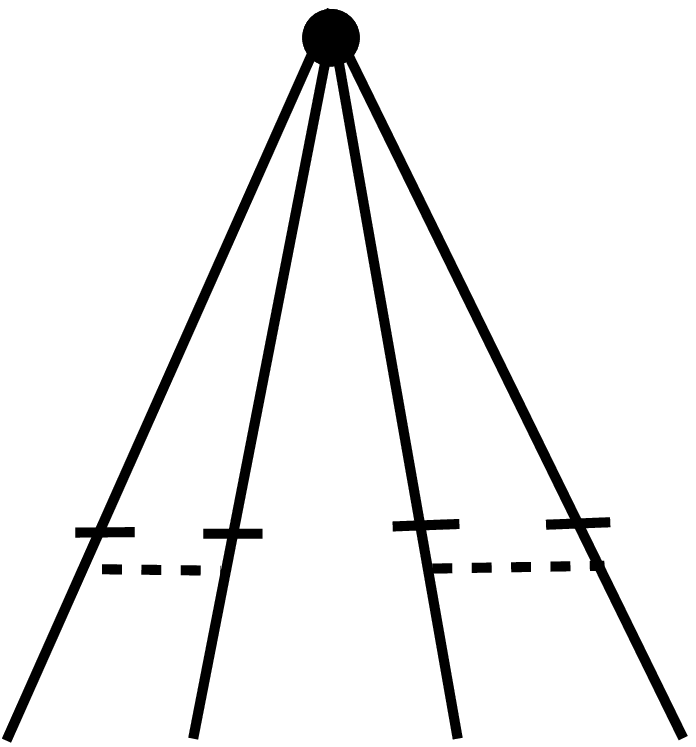}} \\
&
$(D_1)$  &
&
&
$(D_2)$  &
&
$(D_3)$  &
&
$(D_4)$  \\
\end{tabular}
$$

The contribution of a specific diagram into the functional $\Gamma_{nl}$
has the form
\begin{equation}
\label{GammaFinal-scalar}
\bar{\Gamma}=\sum_ip_iA_i,
\end{equation}
where $p_i$ are known combinatorial coefficients and $A_{i}$ are certain
scalar quantities \cite{A_i-----B_i}, \cite{Adzhemyan-Antonov-Honkonen-Kim}.


\section{Calculation of Anomalous Exponents}

In this section we present the two-loop calculation of the critical
dimensions $\gamma^*_F$ of the composite operators $F_{nl},$ which
determine the anomalous exponents in expression (\ref{F-Asympt-OPE}).
We need to extract from the diagrams only the singular parts that
contain the first-order poles in $\epsilon$.



\subsubsection{Calculation of Diagram $D_2$.}
The diagram $D_2$ is represented by the integral
$$ I=\int_{-\infty}^{+\infty}\frac{d\omega}{2\pi}\int_{-\infty}^{+\infty}\frac{d\omega'}{2\pi}\int_m^{\infty}\frac{d\bf{k}}{(2\pi)^d}\int_m^\infty\frac{d\bf{q}}{(2\pi)^d}\cdot\frac{P_{bj}(\bf{k}+\bf{q})}{-i(\omega+\omega')+({\bf k+q})^2}\cdot
$$
$$
\cdot\frac{P_{ai}(\bf{k}+\bf{q})}{i(\omega+\omega')+({\bf k+q})^2}\cdot\frac{P_{em}(\bf{k})}{-i\omega+k^2}
\cdot\frac{P_{ck}(\bf{k})}{i\omega+k^2}\cdot\frac{P_{op}(\bf{q})}{q^{d+\epsilon}}\cdot\frac{P_{qr}(\bf{k})}{k^{d+\epsilon}}\cdot
$$
$$
\cdot\left(i(k+q)_o\delta_{ic}-i(k+q)_c\delta_{io}\right)\cdot\left(i(k+q)_p\delta_{je}-i(k+q)_e\delta_{pj}\right)\cdot
$$
\begin{equation}
\label{D2Raw-calculation}
\cdot\left(ik_q\delta_{kl}-ik_l\delta_{kq}\right)\cdot\left(ik_r\delta_{mf}-ik_f\delta_{rm}\right).
\end{equation}

The use of transversality of the vertex greatly simplifies our calculations:
\begin{equation}
\label{transversality-fieldth}
\mbox{
\begin{tabular}{c c c}
$p_1 V_{123}=p_1(p_2\delta_{13}-p_3\delta_{12})\equiv 0,$ & $\Rightarrow$ & $P_{mn}({\bf p+k})\rightarrow \delta_{nm}.$ \\
\end{tabular}}
\end{equation}

Therefore after contraction of the vector indices with standard symmetric
structures, constructed
from $\delta$-symbols, and differentiation over $m$, which allows to
single out the first-order pole $1/\epsilon$ explicitly, one obtains:
\begin{equation}
\label{D2A1-calculation'}
A_1=\frac{1}{2}\cdot\frac{1}{8}\int_1^\infty dx\int_0^\pi d\theta\left[\frac{x^2+1}{(x^2+2x\cos\theta+1)}-1\right]\frac{\sin^5\theta}{x}=\frac{1}{150};
\end{equation}
\begin{equation}
\label{D2A2-calculation'}
A_2=\frac{1}{2}\cdot\frac{1}{8}\int_1^\infty dx\int_0^\pi d\theta\left[\frac{x^2+1}{(x^2+2x\cos\theta+1)}-1\right]\frac{\sin^3\theta}{x}=\frac{1}{36},
\end{equation}
where $x=q/m.$ All such integrals were calculated analytically for the
most important physical case $d=3$.


\subsubsection{Calculation of Diagram $D_3$.}
In the similar manner, for the diagram $D_3$ one obtains:
\begin{equation}
\label{D3A1-calculation}
A_1=\frac{1}{48}\int_1^\infty dx\int_0^\pi d\theta\left[\frac{x^2+1}{(x^2+x\cos\theta+1)}-1\right]\frac{\sin^5\theta}{x}=\frac{1}{48}\left(-\frac{8\sqrt3}{5}\pi+\frac{656}{75}\right);
\end{equation}
\begin{equation}
\label{D3A2-calculation}
A_2=-\frac{1}{24}\left[\int^{\infty}_{1}dx\int^{\pi}_{0}d\theta\frac{4\cos\theta \sin^3\theta}{x^2+x\cos\theta+1}+\right.
\end{equation}
$$
\left.+\frac{1}{2}\int^{\infty}_{1}dx\int^{\pi}_{0}d\theta\left(\frac{x^2+1}{x^2+x\cos\theta+1}-1\right)\frac{\cos^2\theta \sin^3\theta}{x}\right]=-\frac{1}{24}\left(-\frac{\sqrt{3}}{5}\pi+\frac{24}{25}\right).
$$


\subsubsection{Calculation of Diagram $D_1$.}
For the simplest diagram $D_1$ one obtains
\begin{equation}
\label{D1-calculation}
\mbox{\begin{tabular}{c c}
$A_1=0,$ & $A_2=-\frac{1}{\epsilon}.$
\end{tabular}}
\end{equation}


\subsubsection{Diagram $D_4$.}

The factorized four-ray diagram $D_4$ contains only a second-order pole
in $\epsilon$ and therefore is not needed for the calculation of
$\gamma^*_F$.


\subsubsection{Anomalous Dimension $\gamma_{F_{nl}}^*$.}
The value of anomalous dimension $\gamma_F^*$ is
\begin{equation}
\label{gamma*-calculation}
\gamma_F^*=\sum_i(\bar{p_i}\bar{A_i} u_{*} +
2 \tilde{p_i}\tilde{A_i} u_{*}^2),
\end{equation}
where the quantities with bars and with tildes correspond to the one-loop
and two-loop contributions, respectively.

Finally, combining (\ref{gamma*-calculation}) with (\ref{D2A1-calculation'}),
(\ref{D2A2-calculation'}), (\ref{D3A1-calculation}), (\ref{D3A2-calculation}) and
(\ref{D1-calculation}), for the anomalous dimension of the operator $F_{nl}$
with arbitrary $n$ and $l$ one obtains:
$$\gamma_{F_{nl}}^*=-\left\{\frac{1}{10}\left[n(n+3)-2l(l+1)\right]\cdot\epsilon+\epsilon^2\cdot\left(\frac{2n(n-2)}{125}-\frac{22l(l+1)}{375}+\right.\right.
$$
\vspace{-0.5\baselineskip}
$$+\frac{n(n+3)}{30}+\frac{3}{35}\left(-\frac{\sqrt{3}}{5}\pi+\frac{82}{75}\right)(n-2)\left[2n(n-4)+3l(l+1)\right]-
$$
\vspace{-0.5\baselineskip}
\begin{equation}
\label{gammaAns-calculation}
\left.\left.-\frac{9}{140}\left(-\frac{\sqrt{3}}{5}\pi+\frac{24}{25}\right)(n-2)\left[n(n+3)-2l(l+1)\right] \right) \right\}.
\end{equation}

\section{Comparison with the Exact Solution}

The exact solution for the pair correlator of the problem
(\ref{stochastic-mgd}) was derived in \cite{Lanotte-Mazzino};
see also \cite{Antonov-Lanotte-Mazzino} for a more detailed discussion.
In particular, the exponents
$\zeta_0$ and $\zeta_2$, describing the scaling behavior in the isotropic
and leading anisotropic sectors, were derived exactly for any $d$.
Expanding those expressions to the second order in $\epsilon$ and
setting $d=3$ gives
\begin{equation}
\label{Ep0Dim3-calculation}
\zeta_0=-\epsilon-\frac{1}{3}\epsilon^2, \quad
\zeta_2=\frac{1}{5}\epsilon+\frac{7}{375}\epsilon^2.
\end{equation}
In the RG approach, these exponents should be identified with the
anomalous dimensions of the operators
$\theta_i\theta_i$ and $\theta_i\theta_j$, that is, with
$\gamma^*_{F_{20}}$ and $\gamma^*_{F_{22}}$. It is easily checked that
our expression (\ref{gammaAns-calculation}) is in agreement
with (\ref{Ep0Dim3-calculation}).


\section{Conclusion}

We have applied the RG and OPE methods to the simple
Kazantsev--Kraichnan model, which describes the advection
of a passive vector by the Gaussian velocity field, decorrelated
in time and self-similar in space.

We have shown that the correlation functions of the vector field in the
convective range exhibit anomalous scaling behavior,
what is closely related with existence in this model of composite operators
with negative dimensions. The corresponding anomalous exponents have been
calculated to the second order of the $\epsilon$-expansion (the two-loop
approximation).

It is worth noting that the hierarchy relations between the anisotropic
exponents \cite{Antonov-Lanotte-Mazzino} persist in the two-loop
contributions. It is also worth noting that, in contrast to the scalar case,
the two-loop contributions for scalar operators have the same sign as the
first-order ones, see e.g. (\ref{Ep0Dim3-calculation}) for
$\gamma^*_{F_{20}} = \zeta_{0}$.
Thus the anomalous scaling and the anisotropic hierarchy become even more
strongly pronounced due to the higher-order contributions of the
$\epsilon$-expansion.

The agreement between the exact exponents for the pair correlation function
is also established. This fact strongly supports the applicability of the
RG technique and the $\epsilon$-expansion to the problem of anomalous
scaling for finite values of $\epsilon$, at least for low-order correlation
functions.

\subsubsection*{Acknowledgments.} The authors thank L.Ts. Adzhemyan for
numerous valuable discussions. N.M.G. thanks the Organizers of the
Conference ``Mathematical Modeling and Computational Physics"
(Stara Lesna, Slovakia, July 2011) for the possibility to present
this work.

\end{document}